\documentclass[a4paper,11pt]{article}
\usepackage{pos}
\usepackage{stackengine}
\usepackage{xcolor}
\usepackage{subcaption}
\newsavebox{\panelbox}

\title{Light meson decays at BESIII}

\author*[a]{Kailu Li}
\onbehalf{on behalf of BESIII Collaboration}
\affiliation[a]{Lanzhou University,\\
222 South Tianshui Road, Chengguan District, Lanzhou, China}
\emailAdd{likl@ihep.ac.cn}

\abstract{The world's largest sample of $J/\psi$ events accumulated with the BESIII detector offers a unique opportunity to investigate $\eta$ and $\eta'$ physics via the radiative transition processes $J/\psi \to \gamma\eta/\eta'$. A selection of recent highlights in light meson spectroscopy achieved at BESIII is reviewed in these proceedings, including evidence for the cusp effect in $\eta' \to \pi^0\pi^0\eta$, precision studies of the decay mechanisms of $\eta \to 3\pi$ and $\eta' \to 4\pi$, electromagnetic transition form factor measurements with $\eta/\eta' \to \gamma e^+e^-$ and $\eta' \to \pi^+\pi^-l^+l^-$, as well as searches for rare decays of $\eta$ and $\eta'$ into lepton pairs, including $\eta \to l^+l^-$ and $\eta/\eta' \to l^+l^-l^+l^-$~($l = e, \mu$).}

\FullConference{14th International Workshop on e$^+$e$^-$ collisions from Phi to Psi 2026 (PHIPSI26)\\
8-11 June 2026\\
Pisa, Italy\\}

\begin{document}
\maketitle

\section{Introduction}
The $\eta$ and $\eta'$ mesons are key to low-energy Quantum Chromodynamics~(QCD) studies, offering vital information on mixing parameters, quark masses, and muon $g-2$, while their decays test chiral and vector meson dominance~(VMD) models~\cite{petri} and probe fundamental symmetries, making them a topic of enduring interest in particle physics.

The BESIII detector~\cite{besiii_detector} is a general-purpose spectrometer operating at the Beijing Electron Positron Collider, which is a double-ring $e^{+}e^{-}$ collider running at the center-of-mass energies between 2 and 4.9~GeV. The BESIII experiment has accumulated 10 billion $J/\psi$ events, and copious $\eta$ and $\eta'$ mesons are produced via the radiative and hadronic decays of $J/\psi$. With the radiative decay process $J/\psi \to \gamma\eta/\eta'$, the total $J/\psi$ sample corresponds to $1.1\times 10^{7}$ $\eta$ mesons and $5.2\times 10^{7}$ $\eta'$ mesons, respectively. This large data sample gives BESIII an opportunity to investigate the $\eta/\eta'$ decays with unprecedented statistics.

\section{Evidence for the Cusp Effect in $\eta' \to \pi^{0}\pi^{0}\eta$}
The $S$-wave charge-exchange rescattering $\pi^{+}\pi^{-} \to \pi^{0}\pi^{0}$ produces a cusp at the charged-pion pair threshold, probing low-energy $\pi\pi$ interactions~\cite{meissner,cabibbo,cabibbo_isidori,colangelo,gamiz,bissegger}. Its predicted contribution to $\eta' \to \pi^{0}\pi^{0}\eta$ offers a way to determine $\pi\pi$ scattering lengths~\cite{kubis_schneider,gonzalez}. Earlier measurements found no evidence with the available statistics~\cite{mami,besiii_2018}.

\begin{figure}[htbp]
  \centering
  \begin{minipage}[b]{0.48\textwidth}
    \centering
    \stackinset{r}{15pt}{t}{15pt}{\color{blue}\bfseries (a)}%
      {\includegraphics[width=\linewidth]{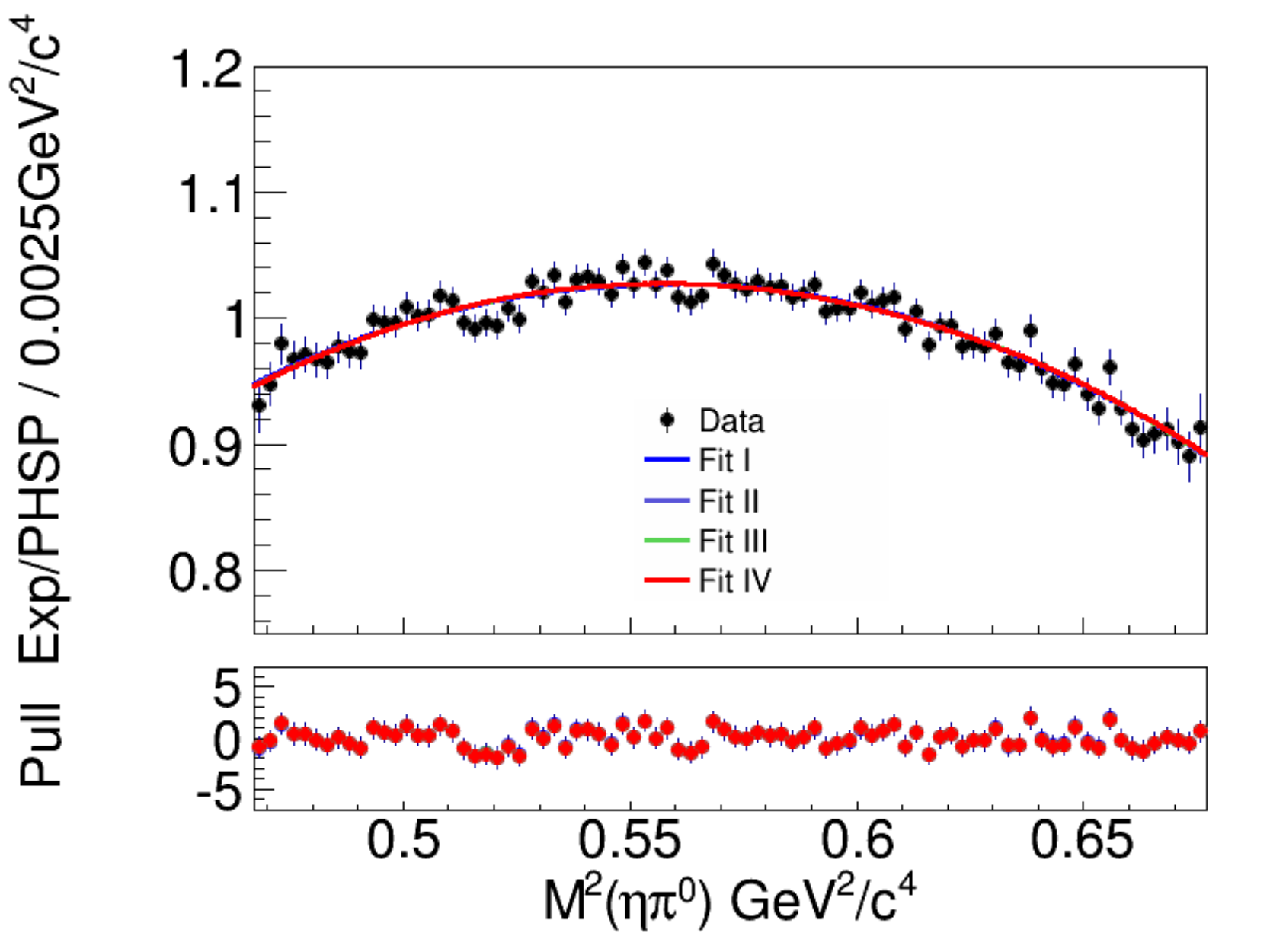}}
  \end{minipage}
  \hfill
  \begin{minipage}[b]{0.48\textwidth}
    \centering
    \stackinset{r}{15pt}{t}{15pt}{\color{blue}\bfseries (b)}%
      {\includegraphics[width=\linewidth]{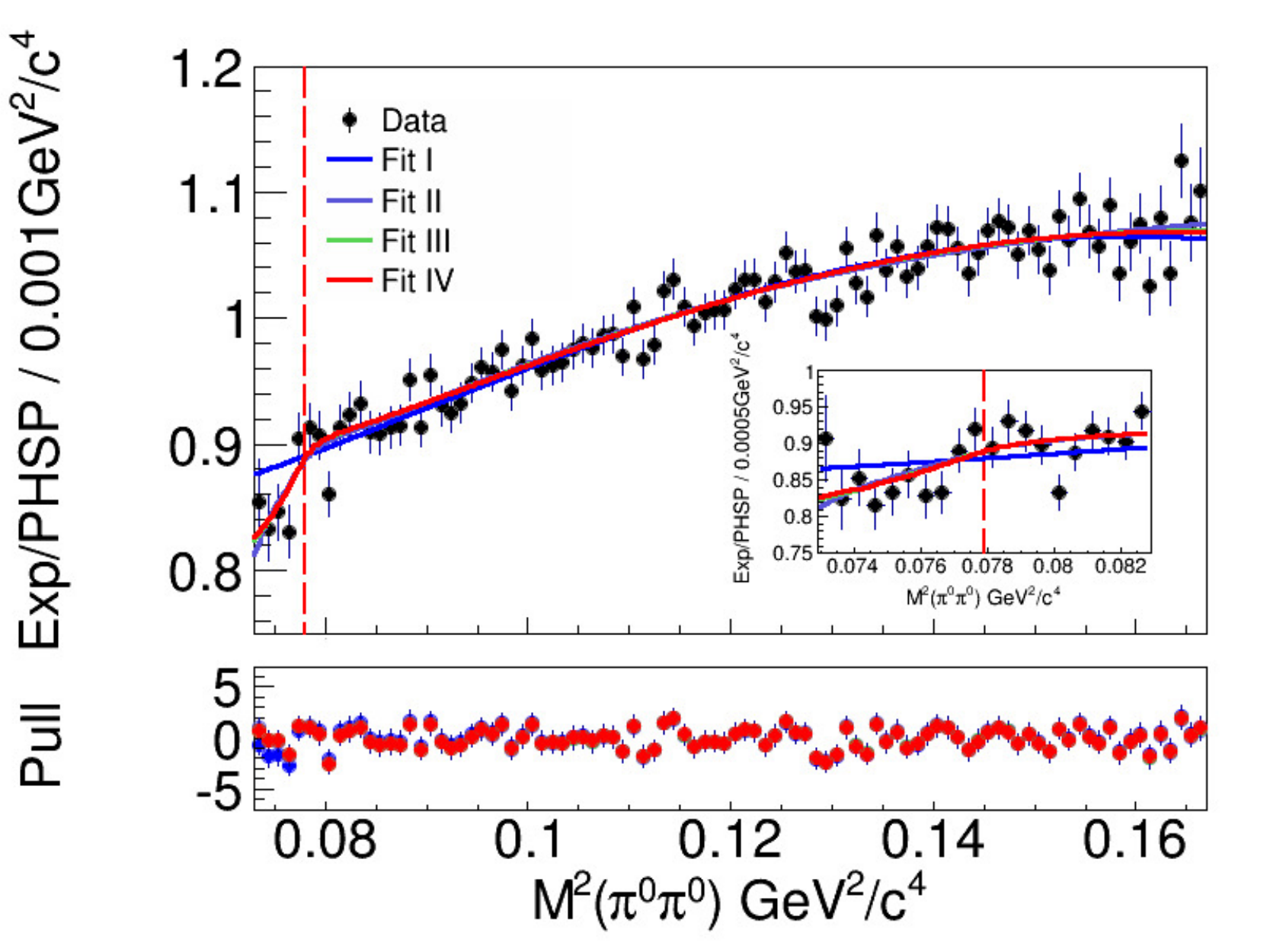}}
  \end{minipage}
  \caption{Comparison of data with fit projections from different models as functions of
    (a) $M^2(\eta\pi^{0})$ and (b) $M^2(\pi^{0}\pi^{0})$, normalized to phase space.}
  \label{fig:cusp}
\end{figure}

With 10 billion $J/\psi$ events collected with the BESIII detector, $4.3\times 10^{5}$ $\eta' \to \pi^{0}\pi^{0}\eta$ events are selected~\cite{cusp_prl}, yielding a sample eight times as large as that previously analyzed by BESIII. A Dalitz plot analysis within the framework of nonrelativistic effective field theory (NREFT)~\cite{kubis_schneider} is performed. A significant discrepancy below the charged-pion pair threshold between data and the tree-level amplitude (Fit~I) is observed, as shown in Fig.~\ref{fig:cusp}, which implies the existence of a sizable contribution from the cusp effect. To describe the data in this region, the one- and two-loop contributions are included in the decay amplitude (Fits~II--IV). The significance of this effect is found to be around $3.5\sigma$, while the combination of $\pi\pi$ scattering lengths $a_0 - a_2$ is determined to be $0.226 \pm 0.060_{\rm stat} \pm 0.013_{\rm syst}$, which is in good agreement with the theoretical calculation~\cite{kubis_schneider}. The evidence for the cusp effect in $\eta' \to \pi^{0}\pi^{0}\eta$ decay demonstrates the potential for investigating the underlying dynamics of light mesons at the BESIII experiment.

\section{Matrix Elements of $\eta \to 3\pi$}
\begin{figure}[htbp]
\centering
\begin{minipage}[b]{0.48\textwidth}
    \begin{picture}(120,150)
      \put(0,0){\includegraphics[width=\linewidth]{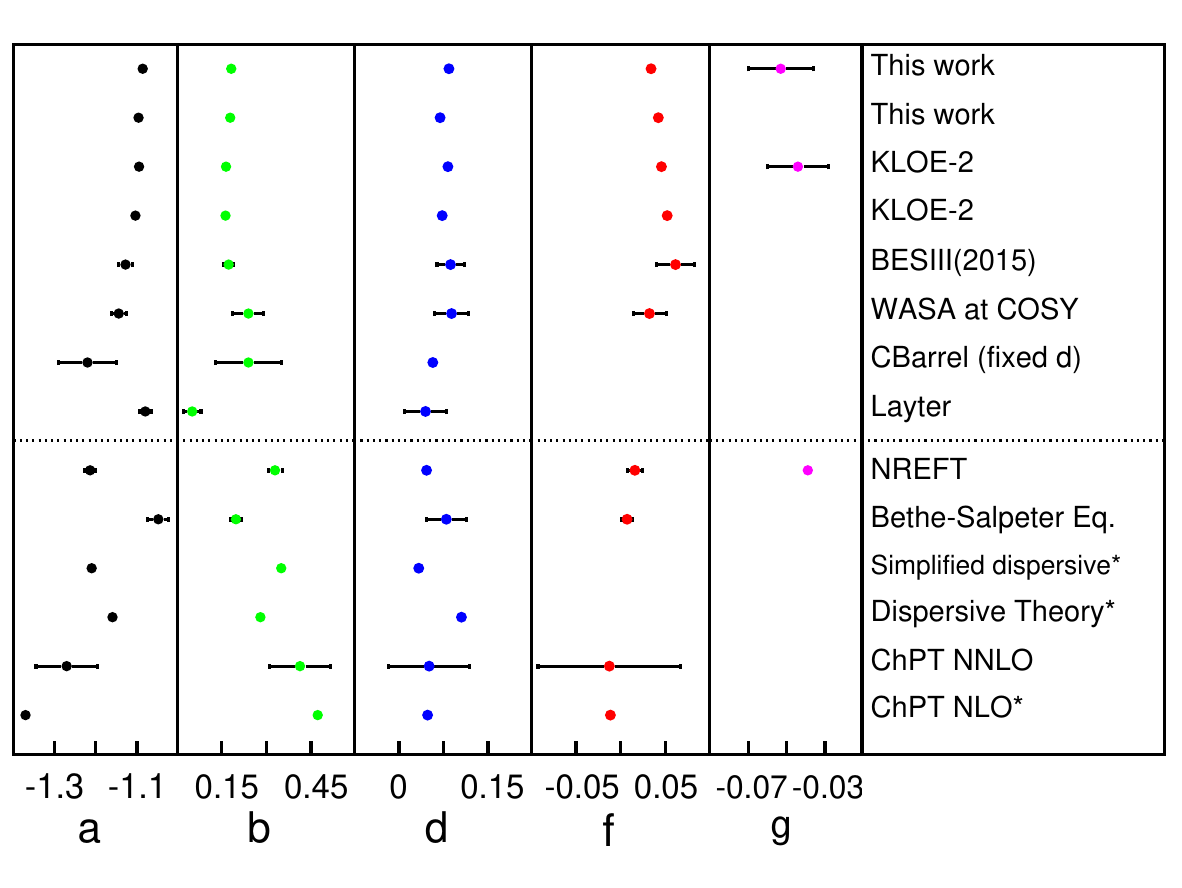}}
      \put(180,130){\color{blue}\scalebox{0.8}{\textbf{(a)}}}
    \end{picture}
\end{minipage}
\hfill
\begin{minipage}[b]{0.48\textwidth}
    \begin{picture}(120,150)
      \put(0,0){\includegraphics[width=\linewidth]{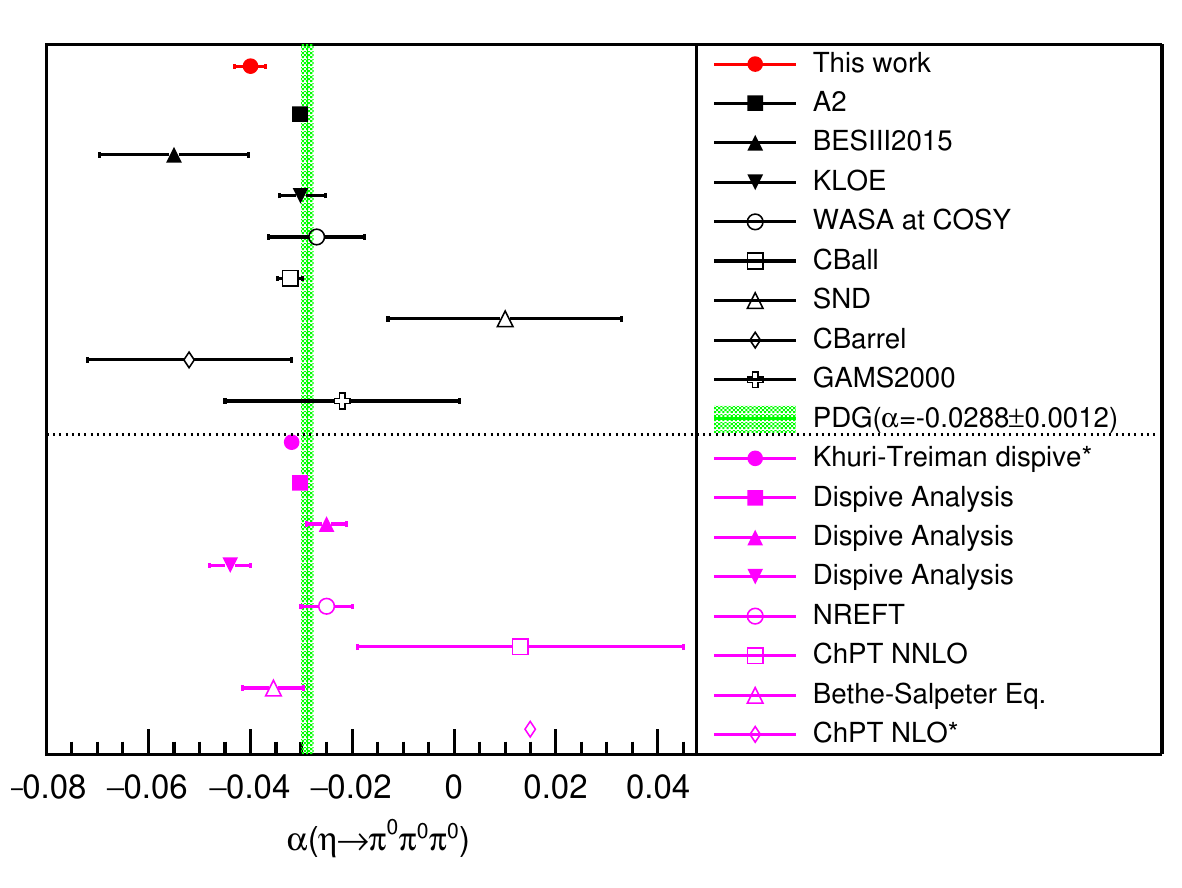}}
      \put(180,130){\color{blue}\scalebox{0.8}{\textbf{(b)}}}
    \end{picture}
\end{minipage}
\caption{Comparison between different experimental results and theoretical predictions for the matrix-element parameters of (a) $\eta \to \pi^{+}\pi^{-}\pi^{0}$ and (b) $\eta \to \pi^{0}\pi^{0}\pi^{0}$.}
\label{fig:3pi}
\end{figure}

The isospin-violating decay $\eta \to 3\pi$ depends on $m_u - m_d$ and constrains the quark mass ratio $Q^2 \equiv (m_s^2 - \hat{m}^2)/(m_d^2 - m_u^2)$, motivating precise matrix-element measurements~\cite{fang_review}.
Using a sample of 10 billion $J/\psi$ decays collected by the BESIII detector, we selected high-purity samples of 631,686 $\eta \to \pi^{+}\pi^{-}\pi^{0}$ events and 272,322 $\eta \to \pi^{0}\pi^{0}\pi^{0}$ events~\cite{eta3pi}. The Dalitz plot for $\eta \to \pi^{+}\pi^{-}\pi^{0}$ is generally described by
\begin{equation}
    X = \frac{\sqrt{3}}{Q_\eta}(T_{\pi^{+}} - T_{\pi^{-}}), Y = \frac{3T_{\pi^{0}}}{Q_\eta} - 1, 
\end{equation}
and the squared amplitude is parameterized as
\begin{equation}
|A(X,Y)|^2 \propto 1 + aY + bY^2 + cX + dX^2 + eXY + fY^3 + gX^2Y + \cdots.
\end{equation}
For the neutral channel, due to the three identical $\pi^{0}$s, the density distribution can be parameterized as
\begin{equation}
|A(X,Y)|^2 \propto 1 + 2\alpha Z + 2\beta(3X^2Y - Y^3) + 2\gamma Z^2 + \cdots,
\end{equation}
with $Z = X^2 + Y^2$.
The charged-channel $C$-violating parameters $c$, $e$, $h$ and $l$ are consistent with zero. A fit retaining $a$, $b$, $d$, $f$ and $g$ gives precise results compatible with recent measurements and theoretical predictions, as shown in Fig.~\ref{fig:3pi}. For the neutral channel, $\alpha = -0.0406 \pm 0.0035 \pm 0.0008$ agrees with the most precise measurement~\cite{a2_mami} within $2.8\sigma$; $\beta$ and $\gamma$ are consistent with zero. A fit including $\pi\pi$ interactions tests the NREFT cusp prediction at the $\pi^{+}\pi^{-}$ threshold~\cite{schneider_rescattering,gullstrom}, yielding $\delta = -0.018 \pm 0.022$ with $0.8\sigma$ significance. No significant cusp is observed.

\section{Chiral Anomalies in $\eta' \to 4\pi$}
The hadronic decays $\eta' \to \pi^{+}\pi^{-}\pi^{+(0)}\pi^{-(0)}$ probe chiral anomalies and the meson-$\gamma^{*}\gamma^{*}$ coupling relevant to the hadronic contribution to the muon anomalous magnetic moment. Their dynamics can be described by VMD through intermediate virtual $\rho$ mesons~\cite{petri}.

BESIII first observed these decays in 2014 using 1.3 billion $J/\psi$ events~\cite{besiii_4pi_2014}, with insufficient statistics for an amplitude analysis. The full 10 billion $J/\psi$ sample yields 1,650 $\eta' \to \pi^{+}\pi^{-}\pi^{+}\pi^{-}$ and 865 $\eta' \to \pi^{+}\pi^{-}\pi^{0}\pi^{0}$ candidates~\cite{besiii_4pi_2024}. Their branching fractions are $(8.56 \pm 0.25_{\rm stat.} \pm 0.23_{\rm syst.}) \times 10^{-4}$ and $(2.12 \pm 0.12_{\rm stat.} \pm 0.10_{\rm syst.}) \times 10^{-4}$, respectively. No significant signal is found for $\eta' \to 4\pi^{0}$, whose $S$-wave is induced by the $CP$-violating QCD $\theta$-term~\cite{pich,ottnad}. The limit is $\mathcal{B}(\eta' \to \pi^{0}\pi^{0}\pi^{0}\pi^{0}) < 1.24 \times 10^{-5}$ at 90\% C.L.~\cite{besiii_4pi_2024}.

\begin{figure}[htbp]
\centering
\begin{minipage}[b]{0.24\textwidth}
    \begin{picture}(120,90)
      \put(0,0){\includegraphics[width=\linewidth]{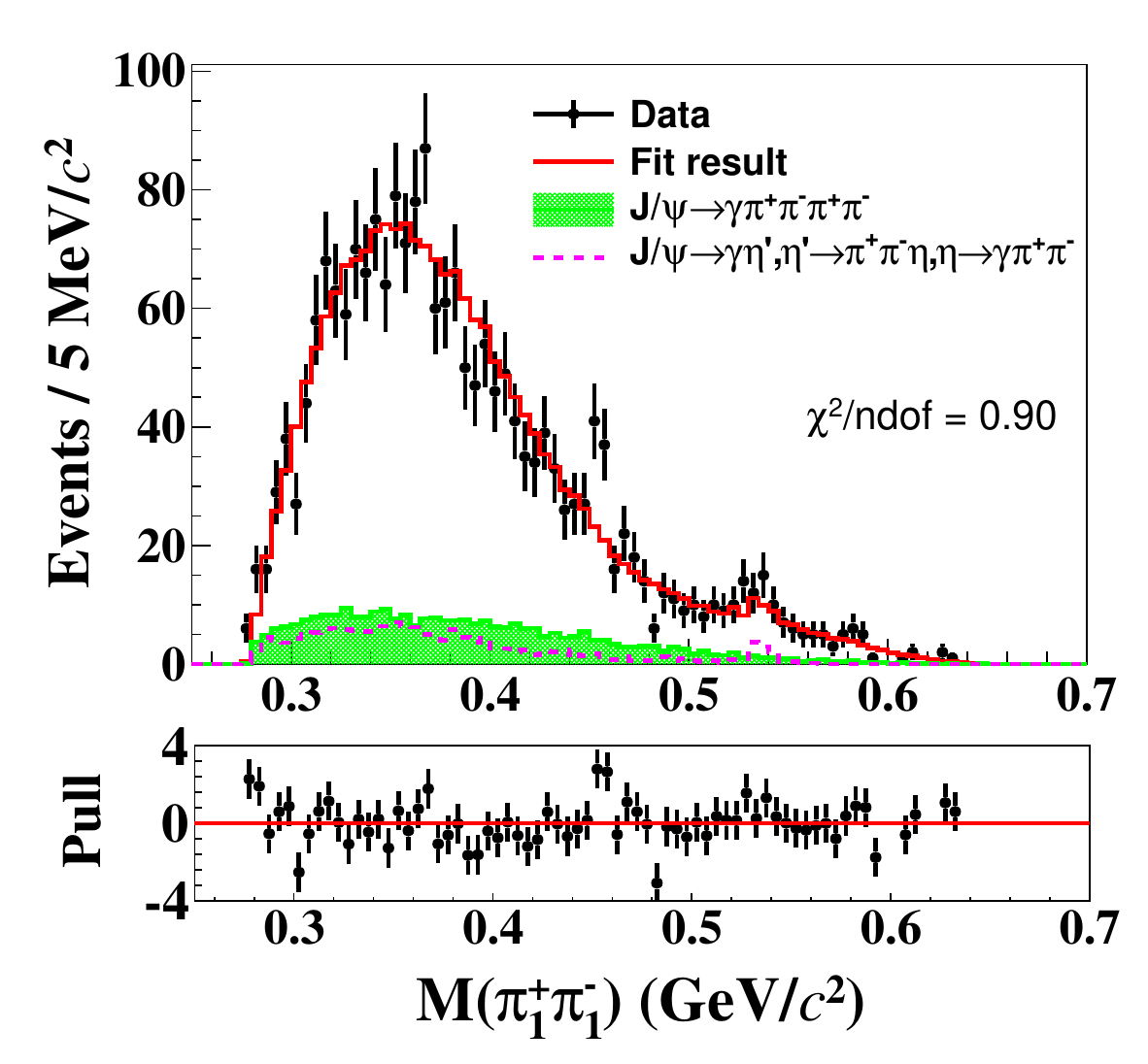}}
      \put(80,80){\color{blue}\scalebox{0.8}{\textbf{(a)}}}
    \end{picture}
\end{minipage}
\begin{minipage}[b]{0.24\textwidth}
    \begin{picture}(120,90)
      \put(0,0){\includegraphics[width=\linewidth]{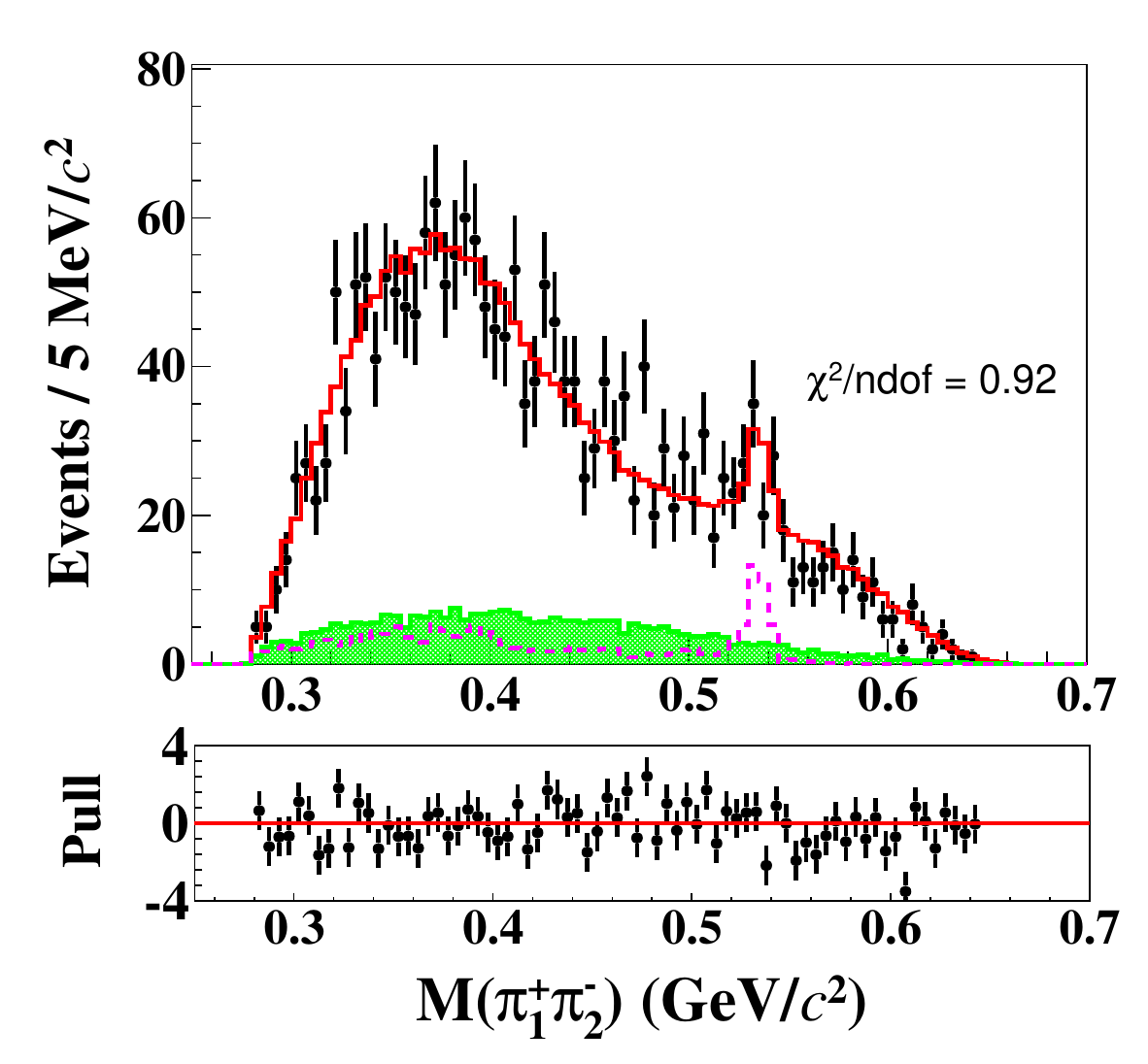}}
      \put(80,80){\color{blue}\scalebox{0.8}{\textbf{(b)}}}
    \end{picture}
\end{minipage}
\begin{minipage}[b]{0.24\textwidth}
    \begin{picture}(120,90)
      \put(0,0){\includegraphics[width=\linewidth]{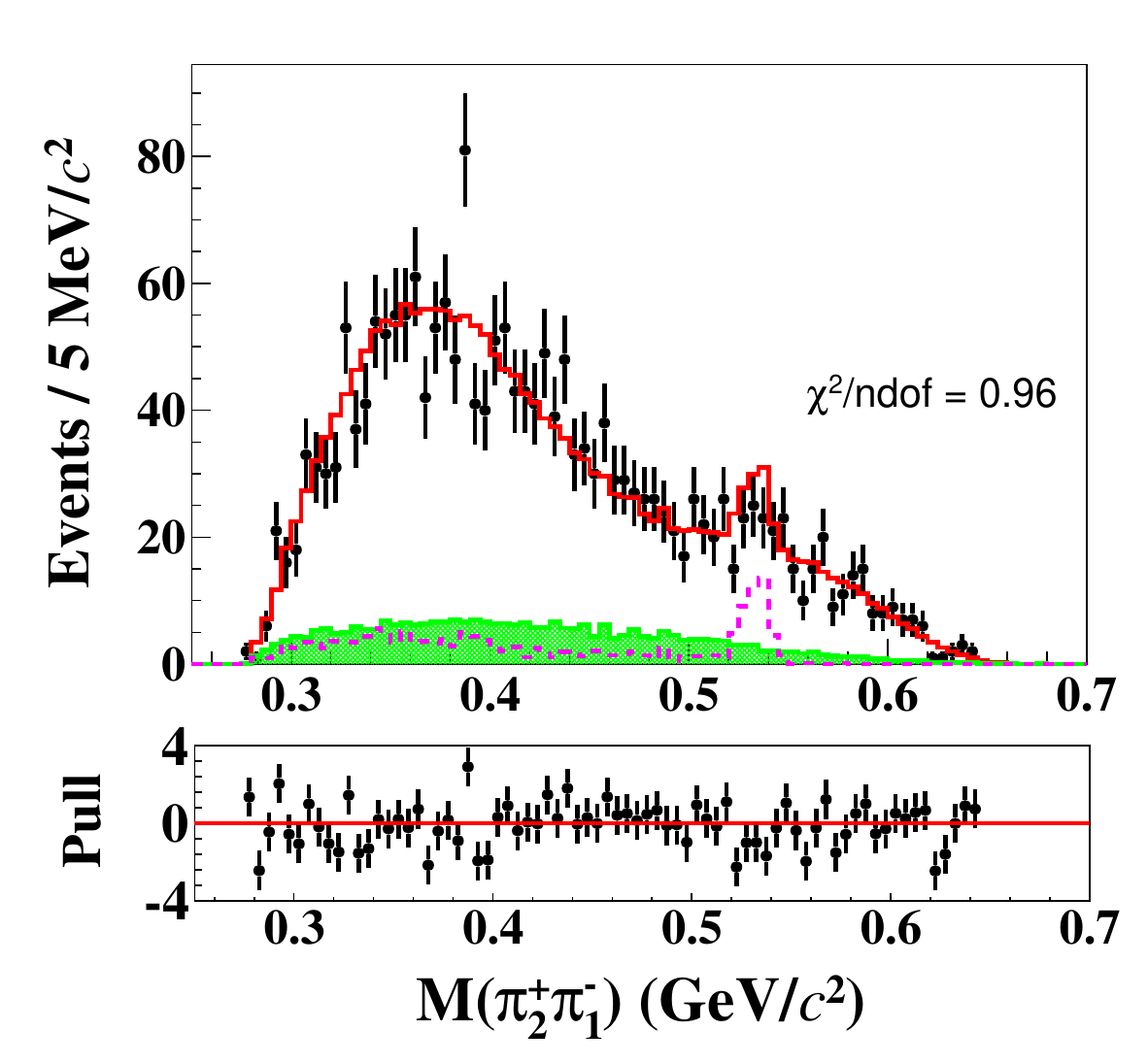}}
      \put(80,80){\color{blue}\scalebox{0.8}{\textbf{(c)}}}
    \end{picture}
\end{minipage}
\begin{minipage}[b]{0.24\textwidth}
    \begin{picture}(120,90)
      \put(0,0){\includegraphics[width=\linewidth]{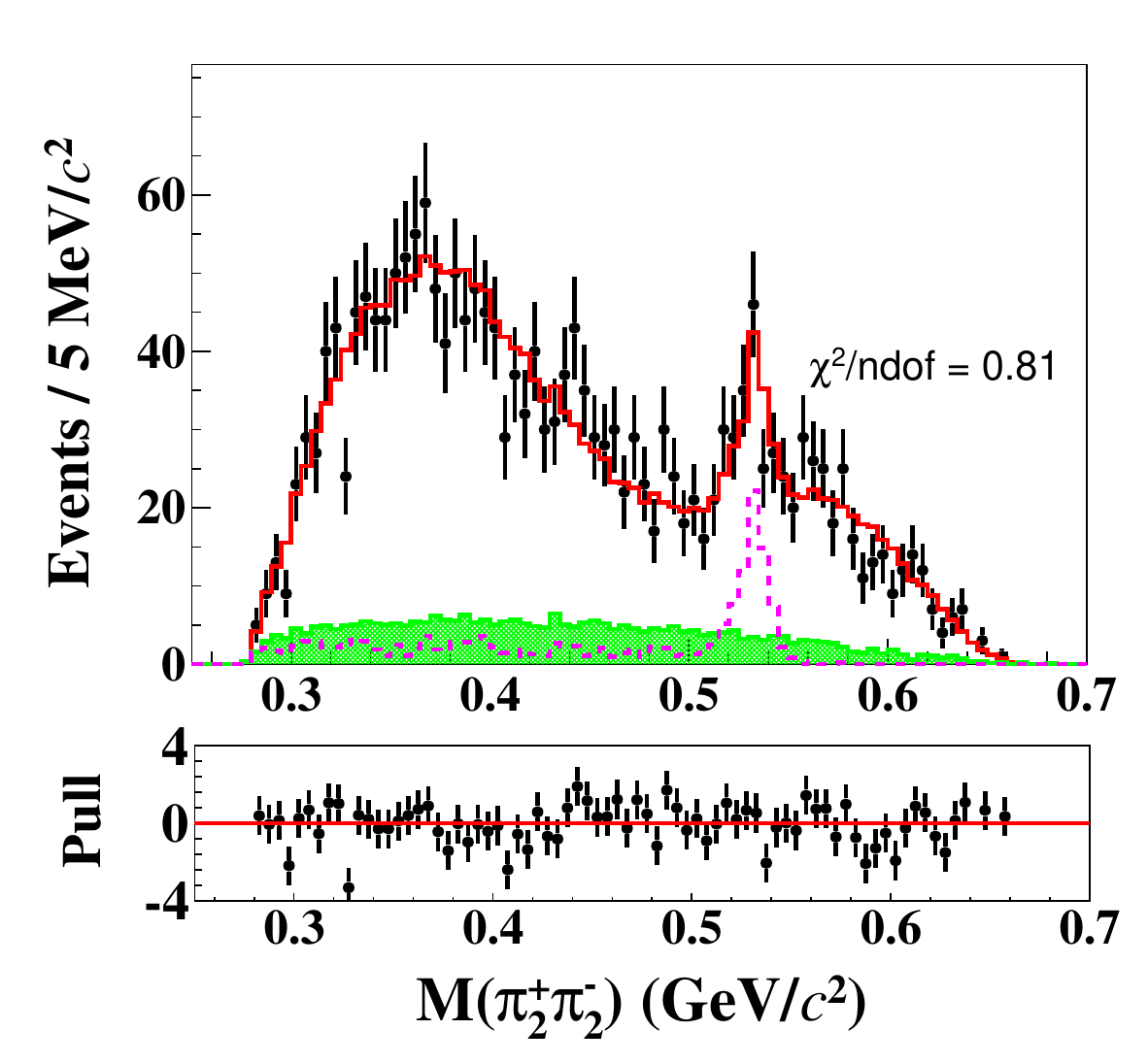}}
      \put(80,80){\color{blue}\scalebox{0.8}{\textbf{(d)}}}
    \end{picture}
\end{minipage}
\caption{Fit projections onto the invariant mass distributions of (a) $\pi^{+}_1\pi^{-}_1$, (b) $\pi^{+}_1\pi^{-}_2$, (c) $\pi^{+}_2\pi^{-}_1$, and (d) $\pi^{+}_2\pi^{-}_2$ in $\eta' \to \pi^{+}\pi^{-}\pi^{+}\pi^{-}$, where $\pi^{+(-)}_1$ and $\pi^{+(-)}_2$ are sorted in ascending order of their energy.}
\label{fig:4pi}
\end{figure}

The charged-channel amplitude includes the box anomaly through one $\rho$ meson and the triangle anomaly through two $\rho$ mesons. The amplitude analysis gives the first determination of the double-virtual isovector form-factor parameter, $\alpha = \frac{c_1 - c_2}{c_3} = 1.22 \pm 0.33_{\rm stat.} \pm 0.04_{\rm syst.}$, consistent with theory~\cite{guo_kubis}. Figure~\ref{fig:4pi} shows the fit projections for the four $\pi^{+}\pi^{-}$ mass combinations.

\section{Form Factors in $\eta/\eta' \to \gamma e^{+}e^{-}$}
Dalitz decays $\eta/\eta' \to \gamma l^{+}l^{-}$ ($l=e,\mu$) proceed through internal conversion of a virtual photon and probe the meson's electromagnetic structure. Deviations from quantum electrodynamics are described by a time-like transition form factor (TFF)~\cite{landsberg}, relevant to the hadronic light-by-light contribution to muon $g-2$~\cite{gm2_wp}. The single-pole form $F(q^2) = 1/(1 - q^2/\Lambda^2)$ describes $\eta$ decays, while $\eta'$ decays use the multi-pole expression $|F(q^2)|^2 = \frac{\Lambda^2(\Lambda^2 + \gamma^2)}{(\Lambda^2 - q^2)^2 + \Lambda^2\gamma^2}$.

\begin{figure}[htbp]
\centering
\begin{minipage}[b]{0.48\textwidth}
    \begin{picture}(120,150)
      \put(0,0){\includegraphics[width=\linewidth]{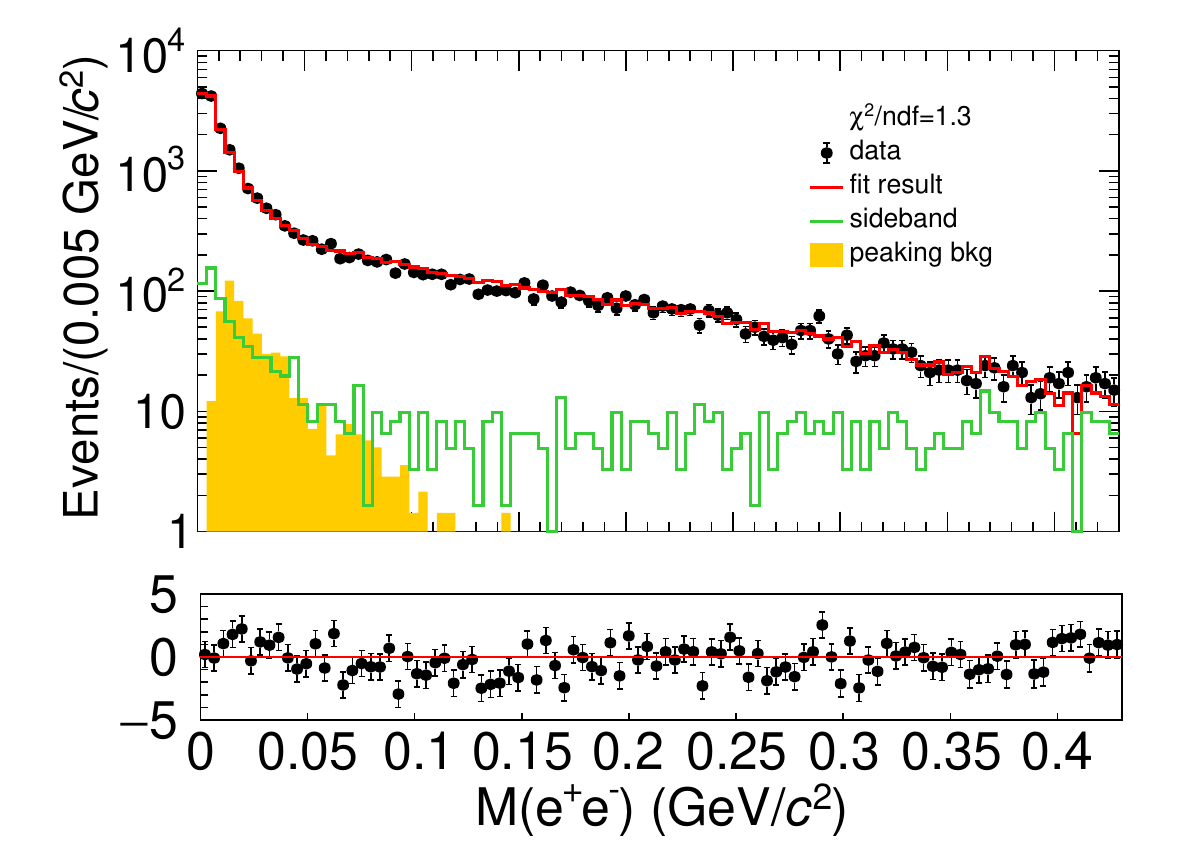}}
      \put(175,125){\color{blue}\scalebox{0.8}{\textbf{(a)}}}
    \end{picture}
\end{minipage}
\hfill
\begin{minipage}[b]{0.48\textwidth}
    \begin{picture}(120,150)
      \put(0,0){\includegraphics[width=\linewidth]{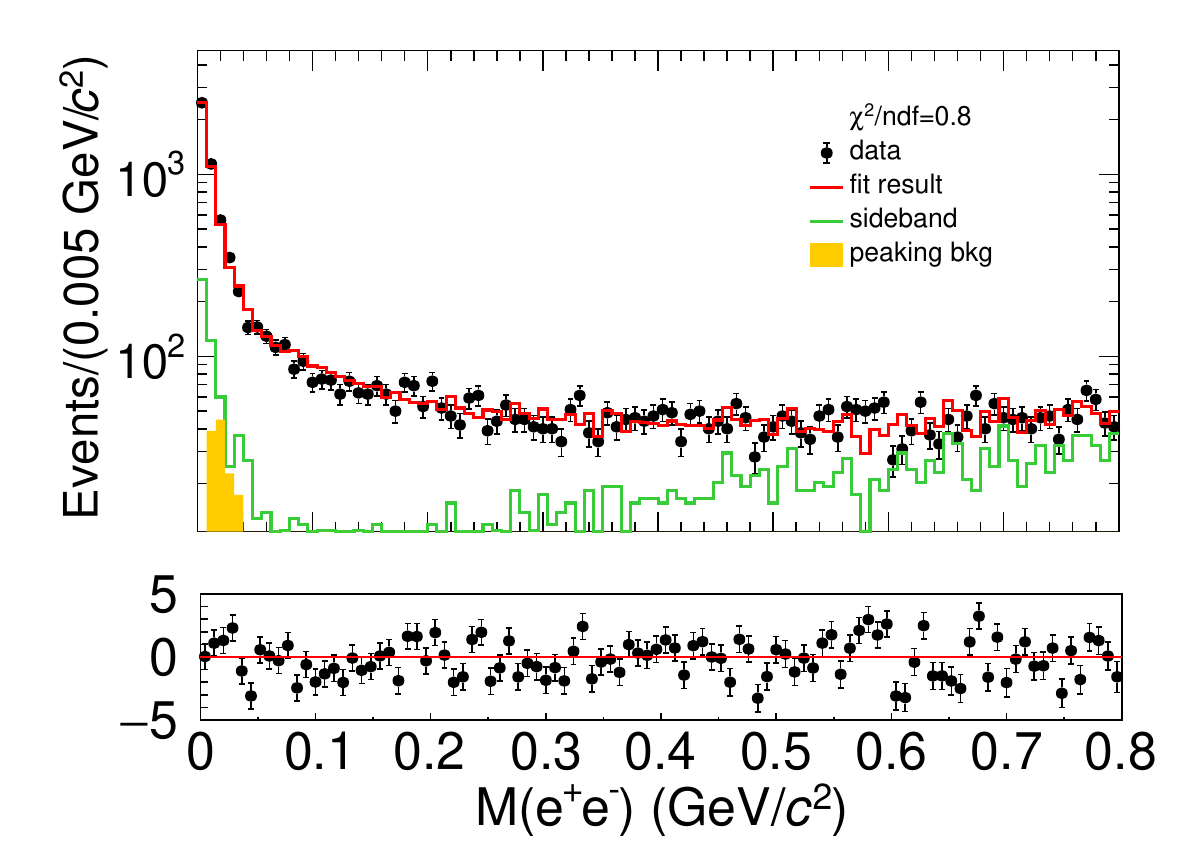}}
      \put(175,125){\color{blue}\scalebox{0.8}{\textbf{(b)}}}
    \end{picture}
\end{minipage}
\caption{Fits to the transition form factors for (a) $\eta \to \gamma e^{+}e^{-}$ and (b) $\eta' \to \gamma e^{+}e^{-}$.}
\label{fig:gamee}
\end{figure}

The 10 billion $J/\psi$ sample yields $2.3\times 10^{4}$ $\eta \to \gamma e^{+}e^{-}$ and $7.6\times 10^{3}$ $\eta' \to \gamma e^{+}e^{-}$ events~\cite{dalitz_ee}. Unbinned maximum-likelihood fits to the $e^{+}e^{-}$ mass spectra (Fig.~\ref{fig:gamee}) give $\Lambda_\eta = (0.749 \pm 0.027 \pm 0.008)$~GeV/$c^2$, $\Lambda_{\eta'} = (0.802 \pm 0.010 \pm 0.002)$~GeV/$c^2$, and $\gamma_{\eta'} = (0.113 \pm 0.010 \pm 0.002)$~GeV/$c^2$, consistent with previous measurements. The interaction-region radii, $R = \sqrt{6b}$, are $(0.645 \pm 0.023 \pm 0.007)$~fm for $\eta$ and $(0.596 \pm 0.005 \pm 0.006)$~fm for $\eta'$.

\section{Precision study of $\eta' \to \pi^{+}\pi^{-}l^{+}l^{-}$}
The decays $\eta' \to \pi^{+}\pi^{-}l^{+}l^{-}$ ($l=e,\mu$) probe the QCD box anomaly and TFFs~\cite{petri}, which enter hadronic corrections to muon $(g-2)_\mu$~\cite{gm2_wp}. Predictions include the hidden gauge and VMD models~\cite{petri} and the chiral unitary approach~\cite{borasoy}. Using 1.31 billion $J/\psi$ events, BESIII first observed $\eta' \to \pi^{+}\pi^{-}\mu^{+}\mu^{-}$ with $8\sigma$ significance~\cite{besiii_ppmm} and precisely measured $\eta' \to \pi^{+}\pi^{-}e^{+}e^{-}$~\cite{besiii_ppee}.

\begin{figure}[htbp]
\centering
\begin{minipage}[b]{0.24\textwidth}
    \begin{picture}(120,90)
      \put(0,0){\includegraphics[width=\linewidth]{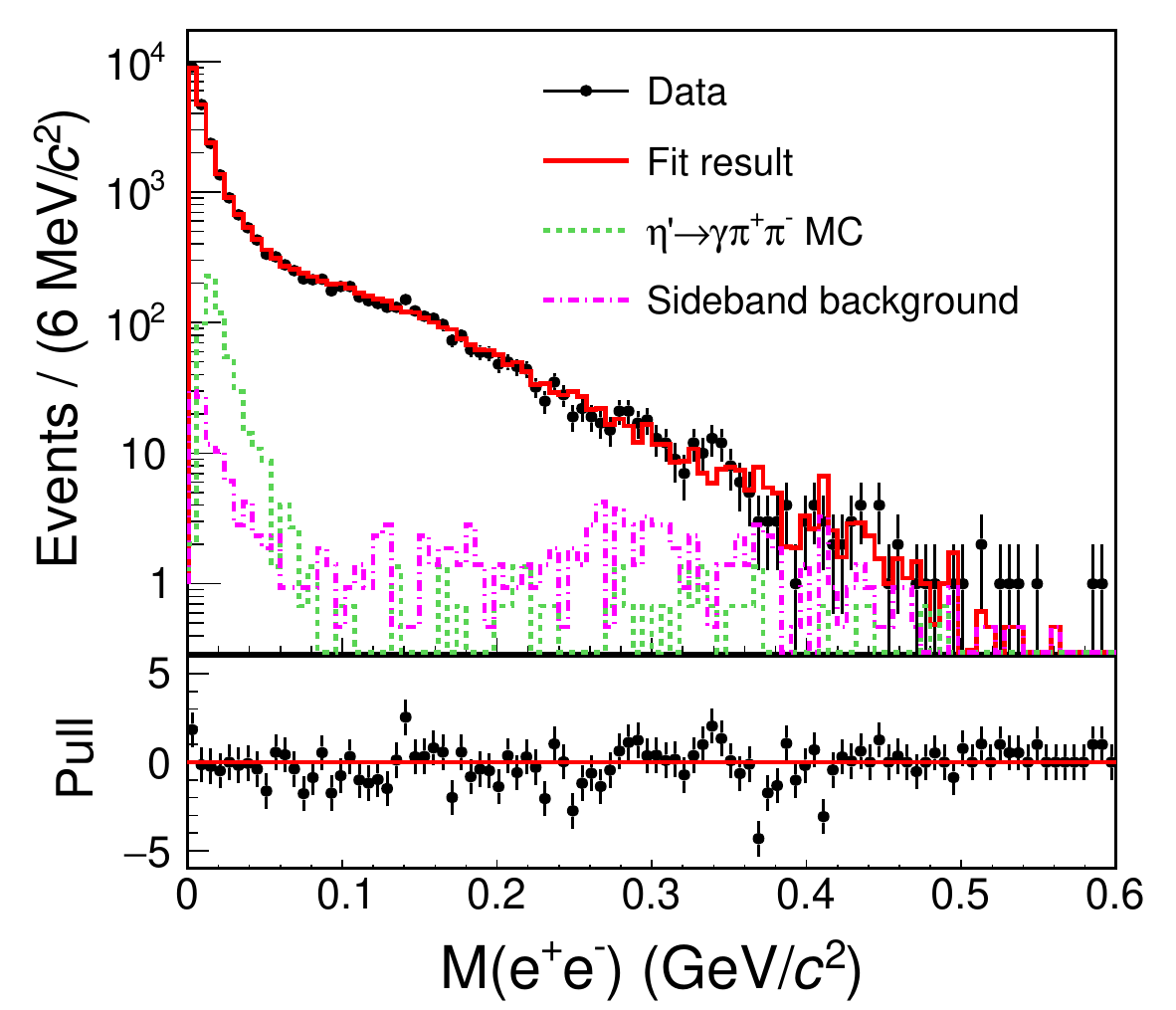}}
      \put(80,78){\color{blue}\scalebox{0.8}{\textbf{(a)}}}
    \end{picture}
\end{minipage}
\begin{minipage}[b]{0.24\textwidth}
    \begin{picture}(120,90)
      \put(0,0){\includegraphics[width=\linewidth]{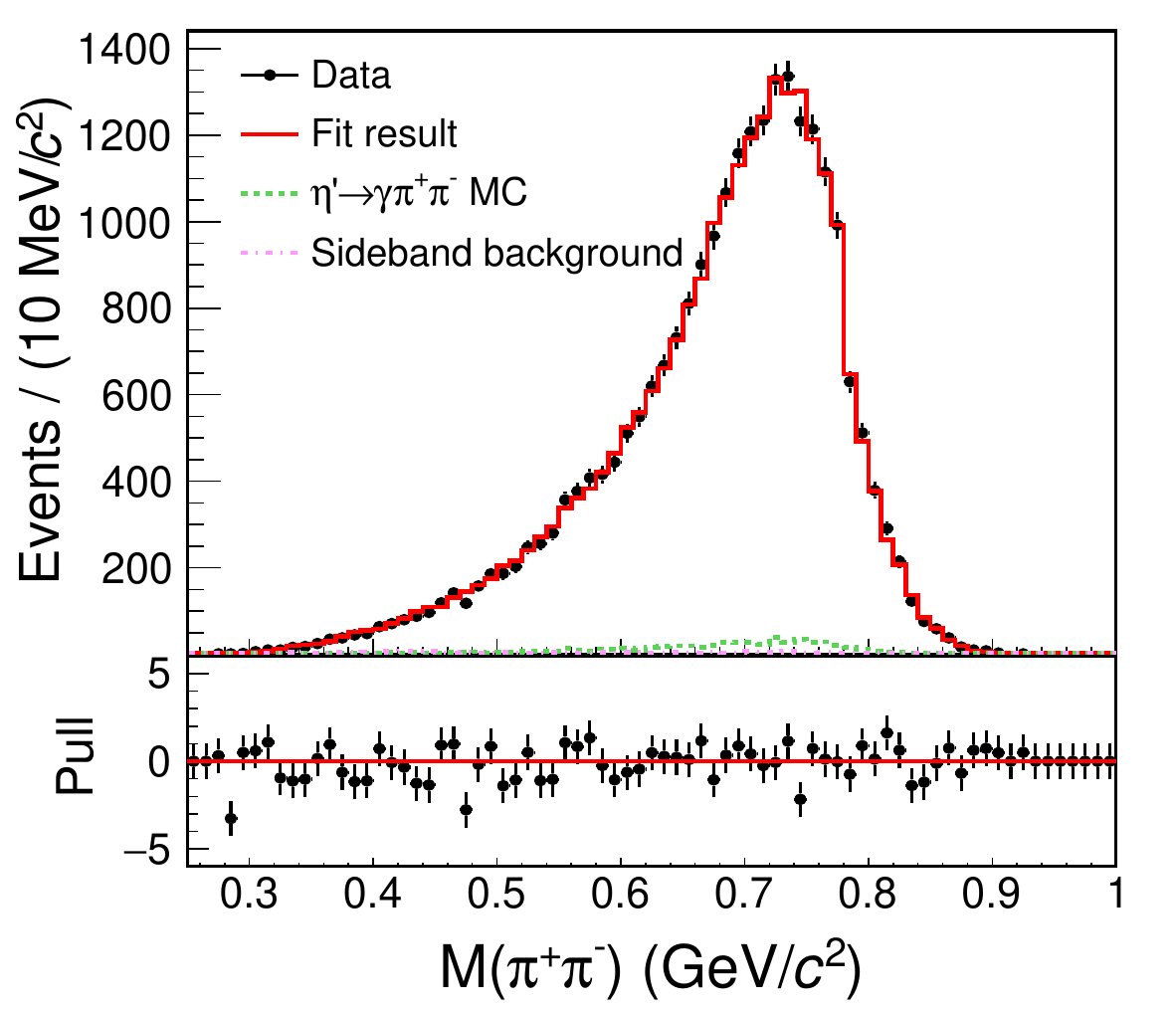}}
      \put(80,78){\color{blue}\scalebox{0.8}{\textbf{(b)}}}
    \end{picture}
\end{minipage}
\begin{minipage}[b]{0.24\textwidth}
    \begin{picture}(120,90)
      \put(0,0){\includegraphics[width=\linewidth]{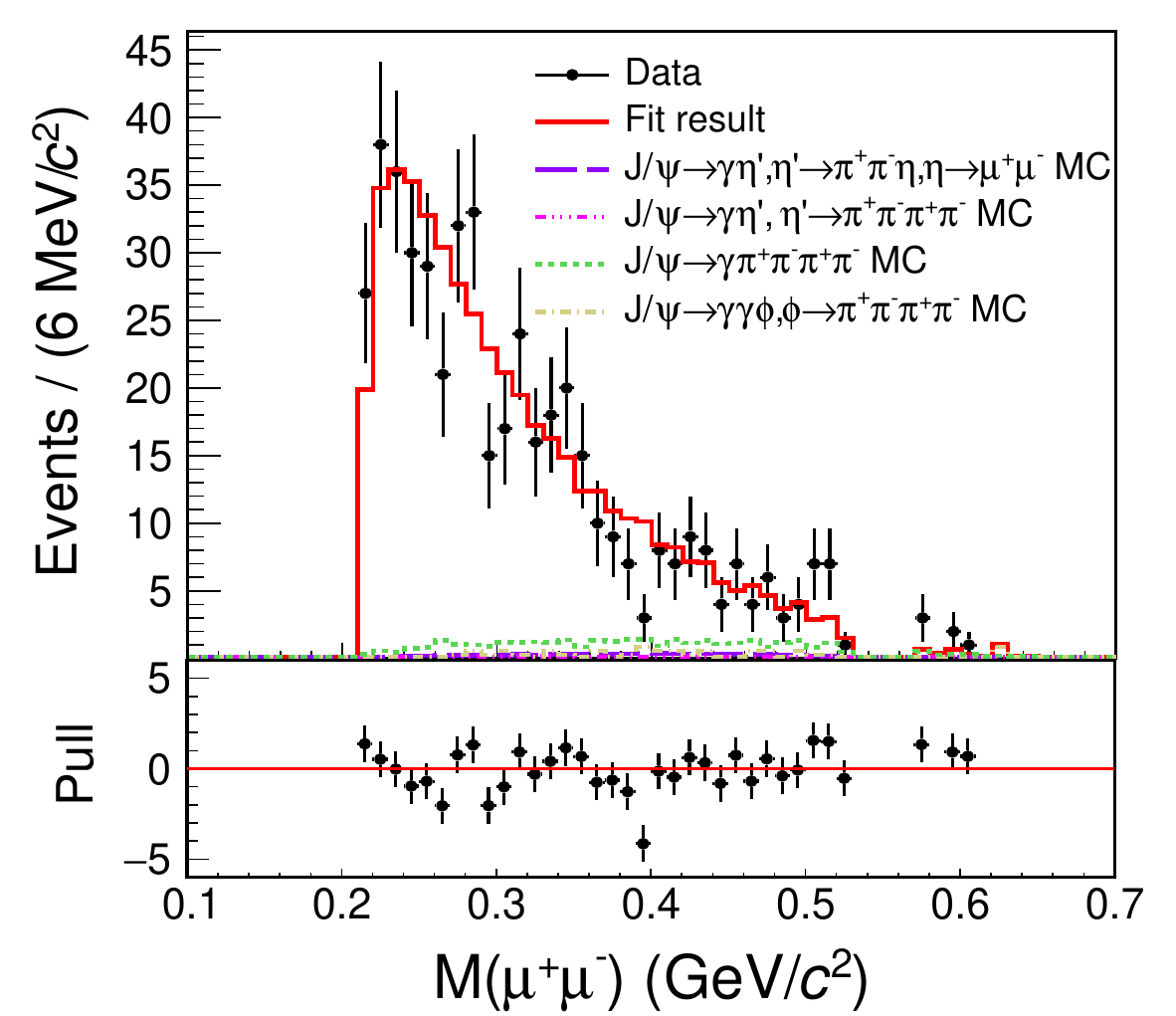}}
      \put(80,78){\color{blue}\scalebox{0.8}{\textbf{(c)}}}
    \end{picture}
\end{minipage}
\begin{minipage}[b]{0.24\textwidth}
    \begin{picture}(120,90)
      \put(0,0){\includegraphics[width=\linewidth]{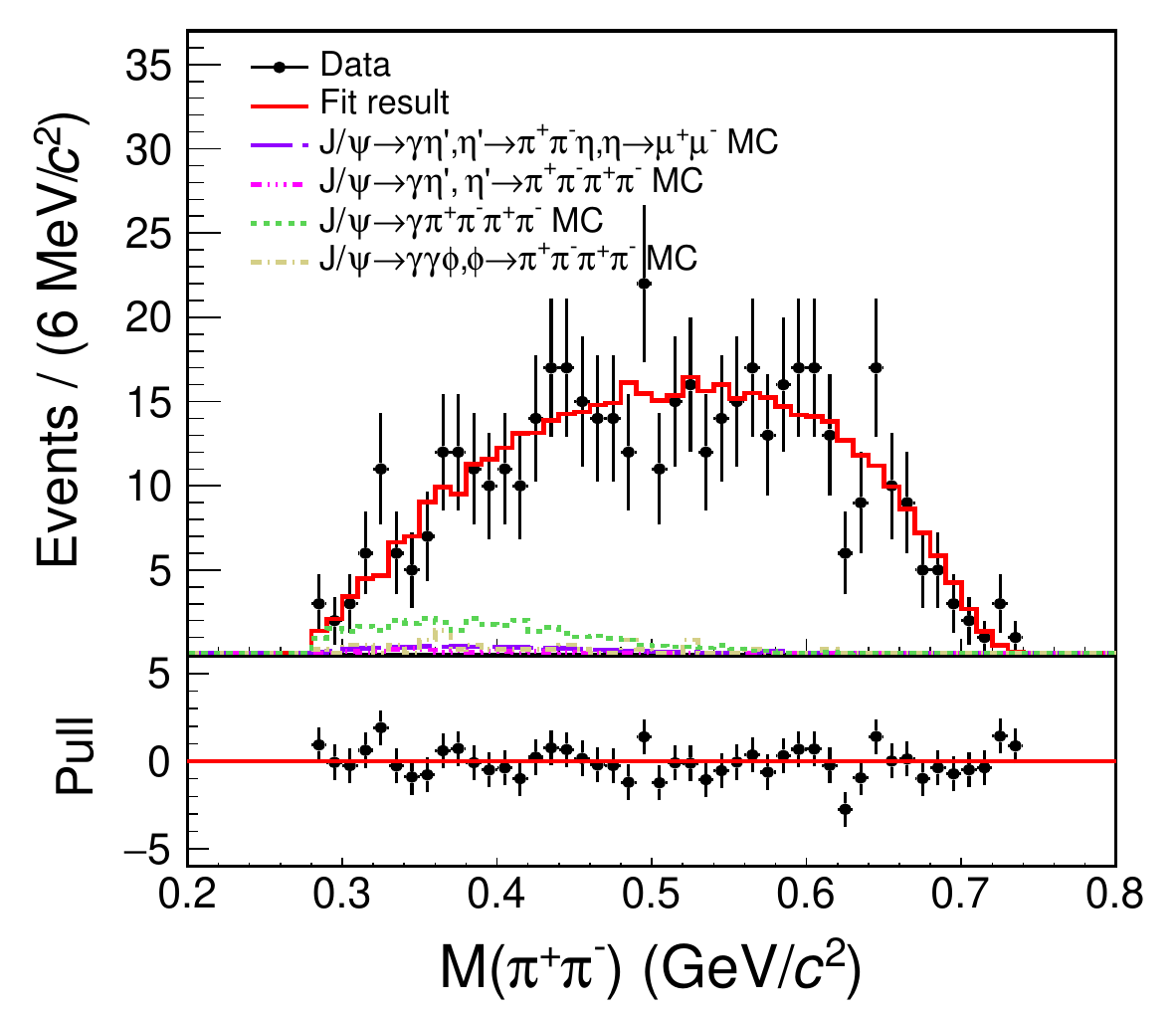}}
      \put(80,78){\color{blue}\scalebox{0.8}{\textbf{(d)}}}
    \end{picture}
\end{minipage}
\caption{Fits to the invariant mass distributions of (a) $e^{+}e^{-}$ and (b) $\pi^{+}\pi^{-}$ for $\eta' \to \pi^{+}\pi^{-}e^{+}e^{-}$, (c) $\mu^{+}\mu^{-}$ and (d) $\pi^{+}\pi^{-}$ for $\eta' \to \pi^{+}\pi^{-}\mu^{+}\mu^{-}$.}
\label{fig:pipill}
\end{figure}

These decays replace the photon in $\eta' \to \gamma\pi^{+}\pi^{-}$ with a virtual photon converting into $l^{+}l^{-}$. Within VMD, the squared amplitude factorizes into axial-anomaly contact terms and VMD contributions with finite-width corrections~\cite{gounaris}. The full 10 billion $J/\psi$ sample provides 22,725 $\eta' \to \pi^{+}\pi^{-}e^{+}e^{-}$ and 434 $\eta' \to \pi^{+}\pi^{-}\mu^{+}\mu^{-}$ candidates~\cite{besiii_ppll}. Unbinned maximum-likelihood fits to the dilepton and dipion mass spectra are shown in Fig.~\ref{fig:pipill}. The electron channel requires an $\omega$ contribution in addition to the $\rho$ to describe the dipion spectrum. The fits agree with the hidden gauge model, but statistical uncertainties prevent discrimination among the tested VMD models.

A combined fit gives the TFF slope $b_{\eta'} = 1.30 \pm 0.19$~(GeV/$c^2$)$^{-2}$, consistent with previous BESIII measurements and VMD predictions~\cite{petri}. This is the first form-factor determination from $\eta' \to \pi^{+}\pi^{-}l^{+}l^{-}$. The angle between the dipion and dilepton decay planes also probes $CP$ violation from an unconventional electric dipole transition~\cite{gao_cp}. The measured asymmetries, $A_{CP}(\eta' \to \pi^{+}\pi^{-}e^{+}e^{-}) = (-0.21 \pm 0.73_{\rm stat.} \pm 0.01_{\rm syst.})\%$ and $A_{CP}(\eta' \to \pi^{+}\pi^{-}\mu^{+}\mu^{-}) = (0.62 \pm 4.71_{\rm stat.} \pm 0.08_{\rm syst.})\%$, show no evidence of $CP$ violation.

\section{Double Dalitz and rare leptonic decays}
Double Dalitz decays $P \to l^{+}l^{-}l'^{+}l'^{-}$, with $P=\pi^{0},\eta,\eta'$ and $l,l'=e,\mu$, proceed through two virtual photons and probe pseudoscalar TFFs. With 10 billion $J/\psi$ events, BESIII measures $\eta/\eta' \to e^{+}e^{-}e^{+}e^{-}$ via $J/\psi \to \gamma\eta/\eta'$~\cite{etapeeee,etaeeee}. The branching fractions and theoretical predictions are compared in Table~\ref{tab:double_dalitz}. For the modes $\eta/\eta' \to l^{+}l^{-}\mu^{+}\mu^{-}$, the signal significances are below $3\sigma$, and upper limits at the 90\% confidence level are set using the Bayesian method~\cite{llmumu}, shown in Table~\ref{tab:double_dalitz}.

The strongly suppressed decays $\eta \to l^{+}l^{-}$ ($l=e,\mu$), dominated by a two-photon intermediate state, are studied in the low-background cascade $J/\psi \to \gamma\eta'$, $\eta' \to \pi^{+}\pi^{-}\eta$~\cite{ll}. A fit to the $\mu^{+}\mu^{-}$ mass spectrum gives $\mathcal{B}(\eta \to \mu^{+}\mu^{-}) = (5.8 \pm 1.0 \pm 0.2) \times 10^{-6}$. No $\eta \to e^{+}e^{-}$ candidates are found in the signal region, yielding $\mathcal{B}(\eta \to e^{+}e^{-}) < 2.2 \times 10^{-7}$ at 90\% confidence level.
%\begin{figure}
%    \centering
%\includegraphics[width=0.33\linewidth, height=0.155\textheight]{figure/4leptons/Fit_etap4e.pdf}
%\includegraphics[width=0.31\linewidth, height=0.16\textheight]{figure/4leptons/Fit_eta4e.pdf}
%\includegraphics[width=0.31\linewidth, height=0.155\textheight]{figure/2leptons/Figure_4.pdf}
%\caption{Fits to the invariant mass distributions of (a) $\eta' \to e^{+}e^{-}e^{+}e^{-}$, (b) $\eta \to e^{+}e^{-}e^{+}e^{-}$ and (c) $\eta \to \mu^{+}\mu^{-}$.}
%\label{fig:4l}
%\end{figure}

\begin{table}[htbp]
\small
\centering
\caption{Theoretical predictions and experimental measurements of the branching fractions for $\eta/\eta' \to l^{+}l^{-}l'^{+}l'^{-}$ decays.}
\label{tab:double_dalitz}
\begin{tabular}{l|ccc}
\hline\hline
Decay mode & \multicolumn{2}{c}{Theoretical predictions} & Experiment \\
\cline{2-4}
 & VMD~\cite{petri} & Data driven~\cite{escribano_dalitz} & BESIII~\cite{etapeeee,etaeeee,llmumu,ll} \\
\hline
$\eta \to e^{+}e^{-}e^{+}e^{-}$ & $2.67(13) \times 10^{-5}$ & $2.71(2) \times 10^{-5}$ & $(2.6 \pm 0.3 \pm 0.2) \times 10^{-5}$ \\
$\eta' \to e^{+}e^{-}e^{+}e^{-}$ & $2.32(4) \times 10^{-6}$ & $2.10(45) \times 10^{-6}$ & $(4.5 \pm 1.0 \pm 0.5) \times 10^{-6}$ \\
$\eta \to e^{+}e^{-}\mu^{+}\mu^{-}$ & $2.15(22) \times 10^{-6}$ & $2.39(7) \times 10^{-6}$ & $< 6.88 \times 10^{-6}$ \\
$\eta' \to e^{+}e^{-}\mu^{+}\mu^{-}$ & $7.97(31) \times 10^{-7}$ & $6.39(91) \times 10^{-7}$ & $< 1.75 \times 10^{-6}$ \\
$\eta \to \mu^{+}\mu^{-}\mu^{+}\mu^{-}$ & $3.80(26) \times 10^{-9}$ & $3.98(15) \times 10^{-9}$ & -- \\
$\eta' \to \mu^{+}\mu^{-}\mu^{+}\mu^{-}$ & $2.19(10) \times 10^{-8}$ & $1.69(36) \times 10^{-8}$ & $< 5.28 \times 10^{-7}$ \\
\hline\hline
\end{tabular}
\end{table}

\section{Summary}
The BESIII experiment has accumulated 10 billion $J/\psi$ events, a unique sample worldwide, allowing studies of light mesons with unprecedented statistics. Based on the $\eta/\eta'$ sample produced via $J/\psi \to \gamma\eta/\eta'$, the BESIII Collaboration has produced fruitful results, including decay dynamics, form factors, discrete symmetry tests, and rare decay searches. For further precision studies, the statistics can be increased by including hadronic $J/\psi$ decays such as $J/\psi \to \phi\eta/\eta'$. Moreover, a novel approach has recently been proposed to investigate $\eta$ decays, in particular its rare or forbidden decays, via $J/\psi \to \gamma\eta', \eta' \to \pi^{+}\pi^{-}\eta$~\cite{novel_approach}. Therefore, more precise measurements of $\eta/\eta'$ decays are expected in the coming years.

\end{document}